\documentclass[aps,prb,a4paper,preprint,showpacs]{revtex4}
\usepackage{amsmath}
\usepackage{amsfonts}
\usepackage{amssymb}
\usepackage{slashbox}
\usepackage{graphicx}
\usepackage{amsbsy}

\begin{document}
\title{Quantum pumping in helimagnet heterostructures}
\author{Rui Zhu\renewcommand{\thefootnote}{*}\footnote{Corresponding author.
Email address:
rzhu@scut.edu.cn; Tel: +86-13556140645; Fax: +8620-87112837.} and Li-Juan Cui}
\address{Department of Physics, South China University of Technology,
Guangzhou 510641, People's Republic of China }

\begin{abstract}

Spin-dependent diffraction occurs in helimagnet-related transport processes. In this work, we investigated quantum pumping properties in the normal-metal/helimagnet/normal-metal heterostructure driven by two out of phase time-dependent gate potentials. At the condition when one of the diffracted beams goes out of the horizon the pumped charge and spin currents demonstrate sharp dips and rises as a function of the helimagnet spiral wavevector $q$. At small and large $q$'s, the transmission and pumping properties approach the behaviors of a ferromagnet and an insulating barrier, respectively. For different helimagnet spiral periods, the diffracted angles are different. As a result, the pumped charge and spin currents demonstrate multiple maximal and minimal peaks as a function of $q$, hence, sensitively depend on the helimagnet spin configuration. All the pumping properties can be interpreted by the quantum gate-switching mechanisms.

\end{abstract}

\pacs {72.10.-d, 75.30.-m, 72.10.Fk}

\maketitle

\narrowtext

\section{Introduction}

 When one or more cyclically slowly varying time-dependent system parameters encircle a finite area in the parameter space with nonzero inside Berry curvature, a dc current can be generated in a mesoscopic conductor at zero bias\cite{Ref1, Ref2, Ref14, Ref5, Ref3, Ref4}. As one of the simplest dynamic processes, quantum pumping serves as a platform to reveal novel properties of newly-discovered quantum states. From the Majorana states\cite{Ref53} to Klein tunneling in graphene\cite{Ref18, Ref19}, from spin-orbit-coupling effect\cite{Ref54} to topological insulating states\cite{Ref35, Ref20}, and etc., the pumped current displays their unique quantum features. Also, spin-dependent pumped current is observed in ferromagnet-involved and spin-orbit-coupling-affected mesoscopic transport systems driven by magnetization precession or gate potential oscillation\cite{Ref54, Ref22, Ref26, Ref55, Ref56, Ref57, Ref58}. However, quantum pumping properties of spatially-nonuniform magnetic states such as the helimagnet was seldom investigated.

 Helimagnet is a kind of stable magnetic structure with spatially-nonuniform magnetization\cite{Ref39, Ref40}. Its spin spirals in space in two or three dimensions. Transmission diffraction occurs in helimagnet-related transport processes\cite{Ref42, Ref43}. A spin-field-effect transistor is proposed based on the two-dimensional electron gas at the surface of multiferroic oxides with a transverse helical magnetic order\cite{Ref51}. Tunneling anisotropic magnetoresistance was observed in the helimagnet tunnel junction\cite{Ref40}. In a conducting helimagnet embedded waveguide, the conductance demonstrates Fano resonance properties, which are shaped by the helimagnet configuration\cite{Ref52}. Besides static transport properties, Tserkovnyak et al. considered the time-dependent magnetic order parameter driven quantum pumping properties in helimagnets and analyzed the evolution of the magnetic spiral\cite{Ref45}. In their studies, the helimagnet spiral varies in time and the sharp diffraction spectrum cannot be seen. In this work, we consider the quantum pumping properties in the normal-metal/helimagnet/normal-metal (NM/HM/NM) heterostructure driven by two out of phase time-dependent gate potentials and investigate the diffraction induced phenomenons. It is found that at the condition when the diffracted beam goes out of the horizon the scattering matrix Berry curvature demonstrates sharp peaks and as a result the pumped current demonstrates multiple dips and rises as a function of the HM spiral wavevector.

\section{Theoretical formulation}

We consider the normal-metal/helimagnet/normal-metal (NM/HM/NM) triple-layer heterostructure modulated by two ac gate potentials depicted in Fig. 1. The Hamiltonians in the NM and HM layers can be formulated respectively as:
\begin{equation}
\begin{array}{l}
{H_{{\rm{NM}}}} =  - \frac{{{\hbar ^2}}}{{2{m_e}}}{\nabla ^2} + {V_1}\left( t \right),\begin{array}{*{20}{c}}
{}&{z < 0,\begin{array}{*{20}{c}}
{}&{z > d,}
\end{array}}
\end{array}\\
{H_{{\rm{HM}}}} =  - \frac{{{\hbar ^2}}}{{2{m^*}}}{\nabla ^2} + J(z){{\bf{n}}_r} \cdot {\bf{\sigma }} + {V_0} + {V_2}\left( t \right),\begin{array}{*{20}{c}}
{}&{0 < z < d,}
\end{array}
\end{array}
\label{eq1}
 \end{equation}
where $d$ is the thickness of the HM
layer. $m_{e}$ and $m^{*}$ are the free and HM multiferroic oxides' effective electron masses respectively. ${\bf{\sigma}}$
is the Pauli vector. $J(z){\bf{n}}_{\bf{r}} $ is the space-dependent
exchange field following the helicity of the HM spiral with ${\bf{n}}_{\bf{r}}  =  \left[ {\sin \theta _r ,0,\cos \theta _r } \right]$,
 $\theta _r = \bar q_m  \cdot {\bf{r}}$, and $\bar q_m  = \left[
{ q,0,0} \right]$.
\begin{equation}
{V_1}\left( t \right) =  - {V_{1\omega }}\cos \left( {\omega t} \right),
\label{eq9}
 \end{equation}
 and
 \begin{equation}
 {V_2}\left( t \right) = {V_{2\omega }}\cos \left( {\omega t + \phi } \right).
  \end{equation}
  $\omega$ is the driving frequency and $\phi$ is the phase difference between the two ac gate potentials. The minus symbol in $V_1 (t)$ is for calculation convenience, which can be absorbed into the Fermi energy $E_F$ and makes no difference to the physical results. From Eq. (\ref{eq1}) it can be seen that the exchange coupling between the electron and
the localized noncollinear magnetic moments within the barrier acts as a nonhomogenous magnetic field. Therefore, spin-dependent diffraction of transmission can be foreseen in the situation.

We consider an ultrathin film of HM with thickness $d=20$ nm, which can be approximated by a Dirac-delta
function. The HM barrier reduces to a plane
barrier. Its Hamiltonian can be rewritten as
\begin{equation}
H_{MF}  =  - \frac{{\hbar ^2 }}{{2m^* }}\nabla ^2  + \left[ {
J{\bf{n}}_r  \cdot {\bf{\sigma }} + V_0 +V_2 (t)} \right]d\delta \left( z
\right),
\end{equation}
where we assume a single spiral layer. $ J =
\left\langle {J\left( z \right)} \right\rangle $ refers to space
and momentum averages of the exchange coupling strength. It should be noted that the helimagnetic
field is sinusoidally space dependent. A multichannel-tunneling
picture should be considered and integer numbers of the helical wave
vector $ q$ could be absorbed or emitted in transmission and
reflection. The two ac gate potentials $V_1 (t)$ and $V_2 (t)$ oscillates slowly so that the adiabatic approximation is justified. To obtain the parameter-dependent scattering matrix, it is sufficient to solve the static Schr\"{o}dinger equation. With a plane wave incidence, the general spinor wave function in the two NM regions
can be written
as
\begin{equation}
\Psi ^{\sigma} \left( {x,y,z} \right) = {e^{i{k_y}y}}\left\{ \begin{array}{l}
{e^{i{k_x}x}}{e^{i{k_z}z}}{\chi _\sigma } + \sum\limits_{\sigma ',n} {r_n^{\sigma \sigma '}{e^{i\left( {{k_x} + nq} \right)x}}{e^{ - ik_{zn}z}}{\chi _{\sigma '}}} ,\begin{array}{*{20}{c}}
{}&{z < 0,}
\end{array}\\
\sum\limits_{\sigma ',n} {t_n^{\sigma \sigma '}{e^{i\left( {{k_x} + n q} \right)x}}{e^{ik_{zn}z}}{\chi _{\sigma '}}} ,\begin{array}{*{20}{c}}
{}&{z > 0,}
\end{array}
\end{array} \right.
\end{equation}
with ${k_z} = \sqrt {2m{E_F}} \cos {\theta _{\texttt{in}}}/\hbar,$  ${k_x} = \sqrt {2m{E_F}} \sin {\theta _{\texttt{in}}}\cos {\phi _{\texttt{in}}}/\hbar,$ ${k_y} = \sqrt {2m{E_F}} \sin {\theta _{\texttt{in}}}\sin {\phi _{\texttt{in}}}/\hbar,$
$\theta_{\texttt{in}}$ and $\phi _{\texttt{in}}$ the incident polar and azimuthal angles, respectively. Here,
$n$ is an integer ranging from $ - \infty $ to $  \infty $ indexing
the diffraction order. And $
k_{zn}  = \sqrt {2m{E_F} - \hbar ^2 k_y^2  - \hbar ^2 (k_{xn}) ^2 }/\hbar $, $k_{xn}  = k_x  + n q$.
No ferromagnet electrode is present. ${\chi _\sigma }$ are the eigenspinors of the ${\sigma _z}$-representation. In the plane ($x$-$y$ plane) perpendicular to the
transport direction ($z$-axis), free motion of the electron is
assumed. Diffraction appears in the $x$-direction and $k_y$ is conserved under translational invariance.

The reflection ($r_n^{\sigma \sigma '}$) and transmission
($t_n^{\sigma \sigma '}$) amplitudes in the $n$th diffraction order
can be numerically obtained from the continuity conditions\cite{Ref40} for $\Psi ^{\sigma}
(x,y,z)$ at $z=0$.
\begin{equation}
\Psi ^\sigma  \left( {x,y,0^ -  } \right) = \Psi ^\sigma
\left( {x,y,0^ +  } \right),
\label{eq6}
\end{equation}
\begin{equation}
\begin{array}{c}
 \frac{{\hbar ^2 }}{{2m_e }}\left. {\frac{{\partial \Psi ^\sigma  \left( {x,y,z} \right)}}{{\partial z}}} \right|_{z = 0^ -  }  + \left[ {V_0 d +V_2 d + {\bf{\tilde w}}\left( {\theta _r } \right)} \right]\Psi ^\sigma  \left( {x,y,0^ -  } \right) \\
  = \frac{{\hbar ^2 }}{{2m_e }}\left. {\frac{{\partial \Psi ^\sigma  \left( {x,y,z} \right)}}{{\partial z}}} \right|_{z = 0^ +  } , \\
 \end{array}
 \label{eq7}
\end{equation}
with
\begin{equation}
{\bf{\tilde w}}\left( {\theta _r } \right) =  Jd\left[
{\begin{array}{*{20}c}
   {\cos \theta _r } & {\sin \theta _r }  \\
   {\sin \theta _r } & { - \cos \theta _r }  \\
\end{array}} \right].
\end{equation}
The continuity equation can be expressed in each
diffracted order and neighboring diffraction orders are correlated by the HM exchange coupling. Transmission amplitudes with arbitrary order cutoff could be recursively obtained by algebra expressed in the following relations.
\begin{equation}
\left[ {\begin{array}{*{20}{c}}
{ - i\tilde V - 2{k_z}}&{\tilde V - 2i{k_z}}\\
{i\tilde V + 2{k_z}}&{\tilde V - 2i{k_z}}
\end{array}} \right]\left( {\begin{array}{*{20}{c}}
{t_0^{uu}}\\
{t_0^{ud}}
\end{array}} \right) = \left( {\begin{array}{*{20}{c}}
{i\tilde J}&{\tilde J}\\
0&0
\end{array}} \right)\left( {\begin{array}{*{20}{c}}
{t_1^{uu}}\\
{t_1^{ud}}
\end{array}} \right) + \left( {\begin{array}{*{20}{c}}
0&0\\
{ - i\tilde J}&{\tilde J}
\end{array}} \right)\left( {\begin{array}{*{20}{c}}
{t_{ - 1}^{uu}}\\
{t_{ - 1}^{ud}}
\end{array}} \right) + \left( {\begin{array}{*{20}{c}}
{ - 2{k_z}}\\
{2{k_z}}
\end{array}} \right),
 \end{equation}
\begin{equation}
\left( {\begin{array}{*{20}{c}}
{ - 2ik_{zn} + \tilde V}&0\\
0&{ - 2ik_{zn} + \tilde V}
\end{array}} \right)\left( {\begin{array}{*{20}{c}}
{t_n^{uu}}\\
{t_n^{ud}}
\end{array}} \right) + \left( {\begin{array}{*{20}{c}}
{\frac{{\tilde J}}{2}}&{\frac{{i\tilde J}}{2}}\\
{\frac{{i\tilde J}}{2}}&{ - \frac{{\tilde J}}{2}}
\end{array}} \right)\left( {\begin{array}{*{20}{c}}
{t_{n - 1}^{uu}}\\
{t_{n - 1}^{ud}}
\end{array}} \right) + \left( {\begin{array}{*{20}{c}}
{\frac{{\tilde J}}{2}}&{ - \frac{{i\tilde J}}{2}}\\
{ - \frac{{i\tilde J}}{2}}&{ - \frac{{\tilde J}}{2}}
\end{array}} \right)\left( {\begin{array}{*{20}{c}}
{t_{n + 1}^{uu}}\\
{t_{n + 1}^{ud}}
\end{array}} \right) = 0.
\label{eq10}
 \end{equation}
 Eq. (\ref{eq10}) is for $n \ne 0$. For convenience of uniform dimension and precise numerical treatment, we change the variables as $\tilde J = {{2{m_e}Jd} \mathord{\left/
 {\vphantom {{2{m_e}Jd} {{\hbar ^2}}}} \right.
 \kern-\nulldelimiterspace} {{\hbar ^2}}}$ and $\tilde V = {{2{m_e}\left( {{V_0} + {V_2}} \right)d} \mathord{\left/
 {\vphantom {{2{m_e}\left( {{V_0} + {V_2}} \right)d} {{\hbar ^2}}}} \right.
 \kern-\nulldelimiterspace} {{\hbar ^2}}}$. Footnotes ``$u$" and ``$d$" indicate the spin-up and down states respectively.
 The opposite spin channels can be treated similarly and the reflection amplitudes are equal to its transmission counterpart by $\delta$-barrier symmetry.
 Transmissivity of a spin-$\sigma$ electron through the HM tunnel
junction with the incident wave vector $[k_x,k_y,k_z]$ to the $n$-th
diffracted order and spin-$\sigma '$ channel with the outgoing wave
vector $[k_{xn},k_y,k_{zn}]$ reads
\begin{equation}
T_n^{\sigma \sigma '} \left( {{E_F},\theta _{\texttt{in}} ,\phi _{\texttt{in}} } \right) =
\frac{{\texttt{Re}\left( {k_{zn} } \right)  }}{{ k_z }}\left|
{t_n^{\sigma \sigma '} } \right|^2.
\label{eq5}
\end{equation}
As all diffracted beams originate from a single incident beam, different diffracted channels coherently contribute to the transport properties with the total transmissivity
\begin{equation}
{T^{\sigma \sigma '}}\left( {{E_F},{\theta _{{\rm{in}}}},{\phi _{{\rm{in}}}}} \right) = {\left| {\sum\limits_{n =  - \infty }^\infty  {\frac{{{\mathop{\rm Re}\nolimits} \left( {{k_{zn}}} \right)}}{{{k_z}}}t_n^{\sigma \sigma '}} } \right|^2}.
\label{eq8}
\end{equation}

We consider quantum pumping driven by two slowly-varying ac gate potentials $V_1 (t)$ and $V_2 (t)$ (see Fig. 1). In this approach, three-order approximation is sufficiently accurate as higher orders decrease exponentially. Numerically diffraction orders of $0$ and $\pm 1$ are included. Soundness of the approximation is further confirmed by the unitarity of the scattering matrix including the $0$ and $\pm 1$ orders.
The transmission amplitudes can be obtained analytically.
\begin{equation}
\begin{array}{c}
r_{ \pm 1}^{ u  d } = t_{ \pm 1}^{ u  d } =\pm i t_{ \pm 1}^{ u  u }=\pm i r_{ \pm 1}^{ u  u }=ir_{ \pm 1}^{ d  d } = it_{ \pm 1}^{ d  d } =  \mp t_{ \pm 1}^{ d  u } =  \mp r_{ \pm 1}^{ d  u }\\
 =  \pm \frac{{\tilde J{k_z}}}{{{{\tilde J}^2} + 4{k_z}{k_{z \pm 1}} + 2i\left( {{k_z} + {k_{z \pm 1}}} \right)\tilde V - {{\tilde V}^2}}},
 \end{array}
 \label{eq2}
\end{equation}
\begin{equation}
\begin{array}{c}
r_0^{ u  d } =t_0^{ u  d }=-r_0^{ d  u } =- t_0^{ d  u }\\
 = \frac{{2i{{\tilde J}^2}{k_z}\left( {{k_{z - 1}} - {k_{z1}}} \right)}}{{\left[ {{{\tilde J}^2} + 4{k_z}{k_{z1}} + 2i\left( {{k_z} + {k_{z1}}} \right)\tilde V - {{\tilde V}^2}} \right]\left[ {{{\tilde J}^2} + 4{k_z}{k_{z - 1}} + 2i\left( {{k_z} + {k_{z - 1}}} \right)\tilde V - {{\tilde V}^2}} \right]}},
  \end{array}
\end{equation}
\begin{equation}
\begin{array}{c}
t_0^{ u  u } = r_0^{ u  u } + 1 =t_0^{ d  d } = r_0^{ d  d } + 1\\
= \frac{{2{k_z}\left[ {{{\tilde J}^2}\left( {{k_{z1}} + {k_{z - 1}} + i\tilde V} \right) - \left( {2{k_z} + i\tilde V} \right)\left( { - 2i{k_{z1}} + \tilde V} \right)\left( { - 2i{k_{z - 1}} + \tilde V} \right)} \right]}}{{\left[ {{{\tilde J}^2} + 4{k_z}{k_{z1}} + 2i\left( {{k_z} + {k_{z1}}} \right)\tilde V - {{\tilde V}^2}} \right]\left[ {{{\tilde J}^2} + 4{k_z}{k_{z - 1}} + 2i\left( {{k_z} + {k_{z - 1}}} \right)\tilde V - {{\tilde V}^2}} \right]}}.
 \end{array}
 \label{eq3}
\end{equation}
The scattering matrix from an incident beam into the $n$-th diffracted one can be expressed as
\begin{equation}
\left( {\begin{array}{*{20}{c}}
{{b_{Ln u }}}\\
{{b_{Ln d }}}\\
{{b_{Rn u }}}\\
{{b_{Rn d }}}
\end{array}} \right) = \left( {\begin{array}{*{20}{c}}
{r_n^{ u  u }}&{r_n^{ d  u }}&{{t'}_n^{ u  u }}&{{t'}_n^{ d  u }}\\
{r_n^{ u  d }}&{r_n^{ d  d }}&{{t'}_n^{ u  d }}&{{t'}_n^{ d  d }}\\
{t_n^{ u  u }}&{t_n^{ d  u }}&{{r'}_n^{ u  u }}&{{r'}_n^{ d  u }}\\
{t_n^{ u  d }}&{t_n^{ d  d }}&{{r'}_n^{ u  d }}&{{r'}_n^{ d  d }}
\end{array}} \right)\left( {\begin{array}{*{20}{c}}
{{a_{L0 u }}}\\
{{a_{L0 d }}}\\
{{a_{R0 u }}}\\
{{a_{R0 d }}}
\end{array}} \right) = {\hat S_n}\left( {\begin{array}{*{20}{c}}
{{a_{L0 u }}}\\
{{a_{L0 d }}}\\
{{a_{R0 u }}}\\
{{a_{R0 d }}}
\end{array}} \right).
\end{equation}
The primed elements are the transmission/reflection amplitudes backwards. Footnotes $L$ and $R$ indicates the left and right electrodes respectively. Here $L$ is the lower NM electrode with $z<0$ and $R$ is the upper NM electrode with $z>0$. For the case of a Dirac-delta barrier and non-ferromagnetic electrodes, ${t'}_n^{\sigma \sigma '} = t_n^{\sigma \sigma '}$ and ${r'}_n^{\sigma \sigma '} = r_n^{\sigma \sigma '}$. Symmetry in the scattering matrix elements in Eqs. (\ref{eq2}) to (\ref{eq3}) is also general in transport through a Dirac-delta barrier.
Unitarity of the scattering matrix including the $0$ and $\pm 1$ orders combined can be proved numerically. The adiabatically pumped $2 \times 2$ tensor current through the HM tunnel junction can be calculated by\cite{Ref3, Ref22}
 \begin{equation}
 {\hat I} = \frac{{e\omega }}{{4{\pi ^2}}}\oint {\left[ {{\mathop{\rm Im}\nolimits} {{\left( {\sum\limits_{m,n =  - \infty }^\infty  {\hat S_m^\dag \frac{{\partial {{\hat S}_n}}}{{\partial {V_1}}}} } \right)}_{LL}}d{V_1} + {\mathop{\rm Im}\nolimits} {{\left( {\sum\limits_{m,n =  - \infty }^{ + \infty } {\hat S_m^\dag \frac{{\partial {{\hat S}_n}}}{{\partial {V_2}}}} } \right)}_{LL}}d{V_2}} \right]} k_{\max }^2\sin {\theta _k}d{\theta _k}d{\phi _k},,
 \end{equation}
where ${k_{\max }} = \sqrt {2m_e {E_F}} /\hbar $ and $m$, $n$ are the diffraction orders.
The pumped charge and spin current follows as
 \begin{equation}
 \begin{array}{l}
{I_c} = {{\hat I}_{11}} + {{\hat I}_{22}},\\
I_{sz} = {{\hat I}_{22}} - {{\hat I}_{11}},\\
I_{sx} =  - \left( {{{\hat I}_{12}} + {{\hat I}_{21}}} \right),\\
I_{sy} =  - i\left( {{{\hat I}_{12}} - {{\hat I}_{21}}} \right),
\end{array}
\label{eq4}
\end{equation}
with the spin angular momentum flow defined in $\hbar {\bf{\sigma }} \cdot {\bf{I}}/2$.

\section{Numerical results and interpretations}

We consider quantum pumping driven by two ac gate potentials in the NM/HM/NM heterostructure (see Fig. 1). In numerical calculations, the NM Fermi energy
$E_F$ is set to be $5.5$ eV. Periods of short-period and long-period HMs are $3$-$6$ nm and $18$-$90$ nm, respectively\cite{Ref44}. In our theoretic approach, we tuned the HM periods from zero to infinity. For the limit of large HM periods, the results approach that of a NM/Ferromagnet/NM heterostructure. For the limit of small HM periods, the results are identical to that of a NM/Normal Insulator/NM heterostructure. Most HM exists in insulating Multiferroic oxides. We set the barrier height of the HM oxide plane $V_0 =0.05$ eV and width $d=20$ nm. We consider diffraction orders of $0$ and $\pm 1$ and numerically proved that keeping the three orders is sufficient for all $ q$ values as higher orders decrease exponentially. The driving amplitude $V_{1\omega}=V_{2\omega}=100$ meV, which is much smaller than the Fermi energy and the HM $\delta$-barrier strength. Phase difference between the two ac gate potentials $\phi$ is set to be $\pi /2$ to demonstrate relatively large pumped current. The driving frequency $\omega$ is in the unit of the pumped current, which can be tuned sufficiently low. Validity of the adiabatic approximation is secured.

During transmission, the incident electron with wave
vector $[k_x,k_z]$ would absorb or emit $n q$ from the
HM and be diffracted into tunnels with wave vector
$[k_{xn},k_{zn}]$.
To demonstrate the diffraction properties, transmission as a function of the HM spiral wave vector is shown
in Fig. 2. For a certain HM spiral wave vector $q$, when one or both of the diffracted beams goes out of the horizon, the order 1 or/and -1 transmission vanishes at a certain incident angle. In Eq. (\ref{eq5}), $k_{z\pm 1}$ becomes imaginary for sufficiently large $q$ therefore the transmission coefficients turns to exactly zero. For the incident $k_x>0$, as $q$ increases, the order 1 diffracted beam vanishes first and the order -1 diffracted beam vanishes at a larger $q$. At the particular $q$ when the diffracted beam vanishes the zero order transmission demonstrates sharp peaks due to abrupt fall of the diffraction strength. It could be seen from Eqs. (\ref{eq2}) to (\ref{eq3}) that $T_1^{\sigma \sigma '}$ ($T_{-1}^{\sigma \sigma '}$) is the same for all the spin channels. The difference in spin channels is a phase factor which affects the scattering matrix Berry curvature and pumping properties. For small $q$'s in comparison with the Fermi wave vector, the diffraction angle is extremely small and the slow variation in the transmissivity deceeds the scale resolution of Fig. 2. $J$ is the exchange coupling strength. The exchange coupling effect between the traveling electron spin and the local HM magnetization counteracts the static electric potential barrier measured by $V_0$. Therefore the transmissivity strengthens for larger $J$'s as shown in Fig. 2. For $J$'s
stronger than $V_0$, the antiferromagnetic exchange coupling itself lifts the system energy and decreases the transmissivity.

From the vertical scale of the three panels in Fig. 2, it could be seen that the spin-conserved zero-order transmission dominates all the three orders and all the spin channels. For spin flipped transmission, the diffracted spectra are stronger than the zero-order ones. Transmissivity of the spin-conserved zero-order is more than ten times larger than the spin-flipped $\pm 1$-order transmissivity. And the spin-flipped $\pm 1$-order transmissivity is about ten times larger than the spin-flipped zero-order transmissivity. This could be interpreted by the spin-dependent HM grating effect. In normal diffraction effect, the first diffracted spectrum strength is about one-digit smaller than the zero-order spectrum. In this case, the exponential decrease in diffracted orders agrees with general grating properties. The spatial modulation is in the spin space. As a result, spin-flipped diffraction is enhanced. The transmission of one-way
incident light through sinusoidal gratings is delta-function-like
strict lines. Analogously, direction of transmission of one-way
incident electron through sinusoidal helimagnet is discrete strict
lines of different grating orders and the spin is conserved or
flipped.

When the HM spiral period is infinitely large or the spiral wavevector $q=0$, the HM reduces to a ferromagnet with spins pointing to the $z$-direction in our case. It could be analytically confirmed that the scattering amplitudes with the three diffracted orders combined of the HM heterostructure are exactly the same to a ferromagnet and hence all the transport properties are the same. Actually in this limit, the three diffraction beams converge into one. So the exact sameness holds for arbitrary-order cutoff. To the opposite limit with $q$ approaching infinity, the HM spin points to any direction and the HM physics approaches that of a normal insulator with no exchange coupling effect. In Fig. 3, numerical results of the total transmissivity defined by Eq. (\ref{eq8}) as a function of $q$ are shown. The asymptotic behaviors of large and small $q$'s are obvious. No spin-flipped transmission occurs for ferromagnet and normal insulator barriers, which also justifies the small and large-$q$ asymptotic behaviors of panel (b). The jumps occur at the two diffraction vanishing points discussed above. In the region between the two limits, the transmission demonstrates a two step jump from strong spin polarization to no spin polarization. It could be interpreted by the picture that as $q$ increases, spatial variation of the HM field strengthens and the spacial accumulated effect is the counteraction between the up and down states, which weakens the transmission polarization. For small $q$'s, $T^{uu}$ and $T^{dd}$ increase with $q$ as the grating effect enhances spin-polarized transmission. For larger $q$'s when both the diffracted beams disappear, transmission becomes non-polarized slightly above the insulator case due to remained small grating effect. For even larger $q$'s the transmission in the HM heterostructure reproduces that of a normal insulator.

It can also be seen in panel (b) that as $q$ increases, $x$-component exchange coupling enters and increases giving rise to spin-flipped transmission. Difference between the $+1$ and $-1$ order diffraction is shown in panel (b) of Fig. 2. $T^{ud}$ and $T^{du}$ demonstrates a peak in between the two diffraction vanishing points. For small $q$'s below the $+1$ order diffraction vanishing point, coherence of the two diffracted beams suppresses spin flipped transmission. For large $q$'s above the $-1$ order diffraction vanishing point, the spin flipped transmission sharply decreases to zero due to that the diffracted transmission governs the spin-flipped transport and both of the two diffracted beams go out of the horizon.

Spin-dependent transmission as a function of the exchange coupling strength $J$ for various $q$'s is shown in Fig. 4. The small and large $q$ limits are justified for all $J$'s. In panel (a), it can be seen that as $q$ increases, $T^{uu}$ first increases from the ferromagnet case and then decreases till it finally falls to the insulator case. For weak exchange coupling, difference in the transmission between different $q$'s is small. From Fig. 3 it could be seen that $q=0.1$ \AA$^{-1}$ is far below the first diffraction-vanishing point close the the ferromagnet $q=0$ \AA$^{-1}$ case. $q=1$ is between the two diffraction vanishing points. $q=10$ and 100 \AA$^{-1}$ are far above the diffraction-vanishing points and approach the insulator limit. It could be seen from all panels in Fig. 4 that the transmission at $q=0.1$ \AA$^{-1}$ is the same as that of the ferromagnet to the resolution of the figure. The spin-flipped transmission at $q=10$ and $100$ \AA$^{-1}$ is zero which is the same to both the ferromagnet and the insulator. The spin-conserved transmission at $q=10$ and 100 \AA$^{-1}$ approaches but not equal to the insulator results. Transmission for $q=1$ \AA$^{-1}$ characterizes the HM transport properties. The exchange coupling effect between the traveling
electron spin and the local HM magnetization counteracts the static electric potential barrier. The transmissivity increases with $J$ for small $J$'s. For $J$'s stronger than the static barrier strength, the antiferromagnetic exchange coupling itself lifts the system energy and decreases the transmissivity, which is demonstrated in panels (a) and (b). In panels (c) and (d) for spin down incidence the exchange coupling can be looked as ferromagnetic, it increases the transmission in all the parameter regime. In panels (a) and (d) $T^{uu}$ and $T^{dd}$ are slightly larger than that of an insulator for relatively large $q$'s and $J$'s. This could also be interpreted by Fig. 3. For $q$'s larger than but not far away from the two diffraction points, $T^{uu}=T^{dd}$ and are slightly larger than $T_I$ as a result of the residual grating effect. The larger the exchange coupling, the stronger the HM grating effect. The transmission approaches the insulator limit at a larger spiral vector.

The pumped charge and spin currents in a single transport direction are shown in Fig. 5. It can be seen that corresponding to the transmission probability shown in Figs. 2 to 4, the pumped current demonstrates sharp dips and rises at the HM spiral wavevectors that the diffracted spectrum disappears. As the diffracted spectrum dominates the spin-flipped transmission, the $x$ and $y$-component spin current is zero for $q$ smaller than the first diffraction-disappearing point or larger than the second diffraction-disappearing point with zero Berry curvature. The time-dependent gate potentials are applied to the NM and HM. The latter modulates the static potential barrier hight of the HM without changing the magnetization configuration or the exchange coupling strength. The purpose is to more prominently demonstrate the grating effect and avoid multi-factor complexity. The asymptotic behaviors of the pumped charge and spin currents at small and large $q$'s naturally follows that of the transmissivity.

In our proposed device, $V_1 (t)$ is applied to the two NM electrodes modulating the Fermi energy $E_F$. The minus sign in Eq. (\ref{eq9}) means positive $V_1 (t)$ increases $E_F$ instead of suppressing it. $V_2 (t)$ modulates the HM potential barrier. Positive $V_2 (t)$ increases the HM potential barrier hight and suppresses transmission. The cosinusoidal oscillation phase of $V_2 (t)$ is $\phi = \pi /2$ in advance of $V_1 (t)$. Contrary to the sinusoidal oscillation, this means gate $V_1 (t)$ opens in advance of gate $V_2 (t)$. For a barrier gate, i.e. higher barriers generate smaller transmission, this phase difference gives rise to a positive pumped current\cite{Ref18}. In our situation, the diffraction-vanishing points occur at a larger $q$ for increased Fermi energy. As the two-step jump moves to larger $q$'s, $T^{uu}$ decreases and $T^{dd}$ increases at the same $q$. The effect of $V_2 (t)$ is similar to a gate for $T^{uu}$ and similar to an anti-gate for $T^{dd}$. Therefore the $z$-component spin current is positive and demonstrates sharp peaks at two jumps which reflects the substraction effect of spin-up and down transmission. Sign and variation of the charge current is a little bit subtle. It is the addition effect of spin-up and down transmission. $T^{uu}$ and $T^{dd}$ flows in opposite directions. For small and larger $q$'s, there is a net positive charge current under the normal quantum gate-switching mechanisms of the quantum pump\cite{Ref18}. For $q$'s near the two jumps $I_c$ demonstrates a sharp decrease followed by a sharp increase as the spin-up and down transmissions counteract each other.

The $x$ and $y$ components of the pumped spin currents are equal to zero at small and large $q$ limits, which is justified by ferromagnet and insulator transport properties. From definition of the pumped spin current in Eq. (\ref{eq4}), it can be seen that $I_{sx}$ is governed by the added contribution of the up-down and down-up spin-flipped transmission multiplied by a minus sign. From Fig. 3 we can see that $T^{ud}$ and $T^{du}$ both demonstrate a sharp hump in between the two diffraction vanishing points. Therefore $T^{ud}$ and $T^{du}$ both decrease at the left slope and increase at the right slope when $E_F$ is increased and the hump moves to the larger-$q$ region. $V_2 (t)$ acts as a gate at the left slope and acts as an anti-gate at the right slope. Taking into account the minus sign in Eq. (\ref{eq4}), $I_{sx}$ demonstrates a sharp negative dip at the first diffraction-vanishing point and demonstrates a sharp positive peak at the second diffraction-vanishing point. Also defined in Eq. (\ref{eq4}), $I_{sy}$ is governed by the subtracted contribution of the two spin-flipped transmission. We can also see in Fig. 3 (b) that the jump of $T^{ud}$ at both the left and right slopes of the hump is larger than that of $T^{du}$. Considering the minus sign in $I_{sy}$ of Eq. (\ref{eq4}) and the subtraction, $I_{sy}$ reverses sign with magnitude about one-digit smaller relative to $I_{sx}$.

In real transport measurements, the Fermi reservoir emits electrons with Poissonian distribution and usually a single beam transport cannot be generated or detected. We show in Fig. 6 numerical results of the angular-averaged pumped charge and spin currents, which can be directly compared with experiment. The angular-averaged pumped charge and spin currents include contributions from the two diffraction-vanishing steps in transmission and also contributions from a general background pumped current by channels outside the diffraction regime. From panels (a) and (b), it can be seen that the absolute magnitude of $I_c$ is larger than $I_{sz}$ with $I_{sz}$ bearing much stronger oscillations. These observations can be interpreted by the two types of contributions. The general background charge current is larger than that of $z$-component spin current, which is natural since $I_c$ is much above $I_{sz}$ for a ferromagnet and insulator. The diffraction governed contribution gives rise to much stronger oscillations in angular-averaged $I_{sz}$ than angular-averaged $I_c$ due to the opposite directions between the spin-up and down pumped currents discussed previously. For different incident angles, the corresponding diffraction-vanishing $q$ points are different. As a result, the total $I_c$ and $I_{sz}$ contributed from all incident angles demonstrate multiple maximal and minimal points as a function of $q$. Naturally the ferromagnet and insulator limits of the HM at small and large $q$'s are justified in the experimentally observable pumping properties. $I_c$ and $I_{sz}$ decrease as the HM spiral wavevector $q$ is increased due to less orbital channels contributing to the pumped currents. The angular-averaged $I_{sx}$ and $I_{sy}$ are zero for all $q$ values. It is because that the transmission properties are symmetric around $z$-coordinate and contributions from $\theta _{\rm{in}}>0$ and $\theta _{\rm{in}}<0$ channels to $I_{sx}$ and $I_{sy}$ cancel out.

\section{Conclusions}

In this work, we consider the quantum pumping properties in the NM/HM/NM heterostructure driven by two ac gate potentials. As the spin spirals in the $x$-$z$ plane of the HM, spin-dependent diffraction occurs in the scattering process. In the small and large $q$ limits, the transmissivity and pumping properties demonstrate asymptotic behavior towards that of a ferromagnet and normal insulator, respectively. The pumped charge and spin currents of a single transport angle demonstrate a singular increase and fall at the HM spiral wavevector when the diffracted beam goes out of the horizon. The experimentally measurable angular-averaged pumped charge and $z$-component spin currents demonstrate multiple sharp dips and rises as a function of the HM spiral wavevector. The angular-averaged pumped charge and spin currents include contributions from the diffraction process and also contributions from a general background pumped current by channels outside the diffraction regime. All the pumping properties can be interpreted by the quantum gate-switching mechanisms.

\section{Acknowledgements}

We acknowledge enlightening discussions with Jamal Berakdar and Wen-Ji Deng. This project was supported by the National Natural Science
Foundation of China (No. 11004063) and the Fundamental Research
Funds for the Central Universities, SCUT (No. 2012ZZ0076).

\clearpage

\begin{figure}[h]
\includegraphics[height=10cm, width=16cm]{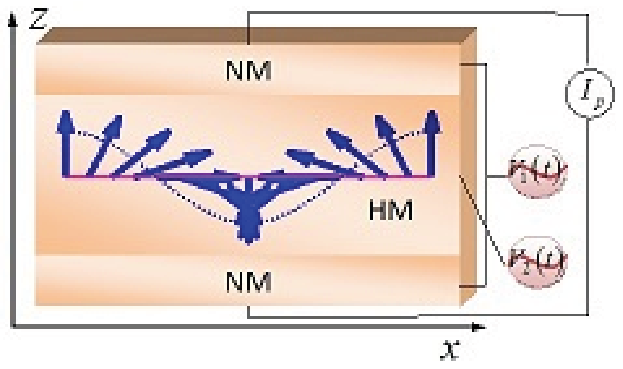}
\caption{Schematics of the quantum pump based on a normal-metal/helimagnet/normal-metal (NM/HM/NM) triple-layer heterostructure. In the HM, the spin spirals in the $x$-$z$ plane in the trigonometric function. The blue arrows and dotted line indicate the spin and the spiral envelope function respectively. The two time-dependent gate potentials are applied to the NM and HM respectively. The HM layer is an extremely thin film. Its thickness is exaggerated to show the helimagnet spiral.}
\end{figure}

\clearpage

\begin{figure}[h]
\includegraphics[height=10cm, width=14cm]{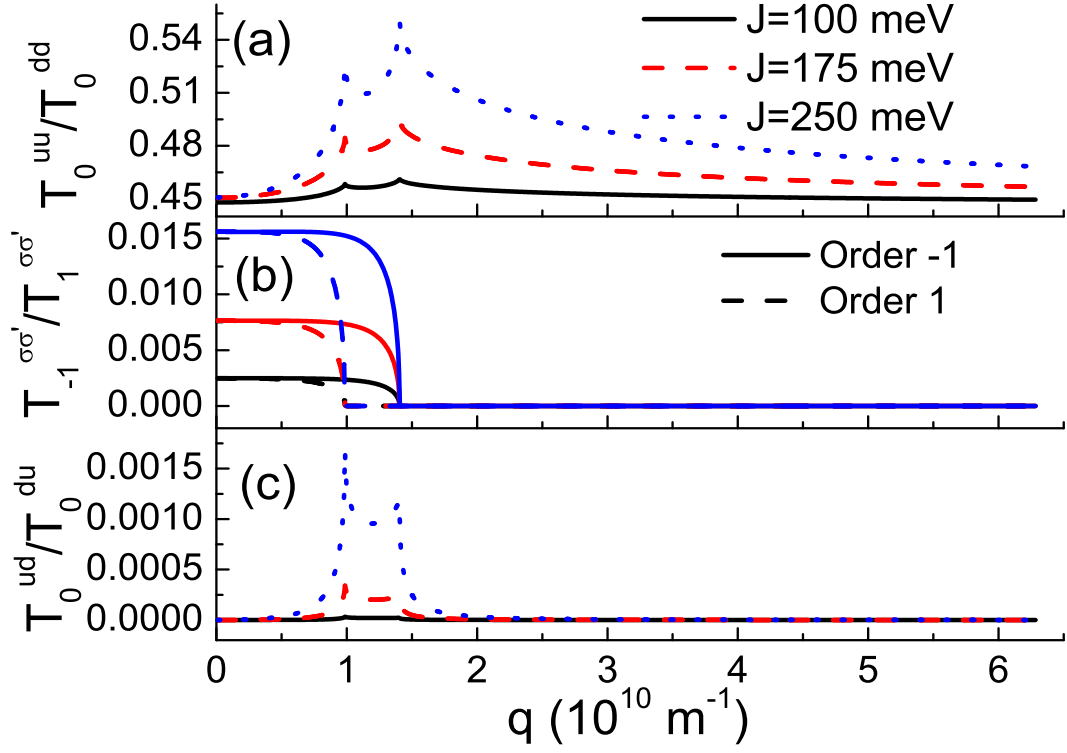}
\caption{Spin-dependent transmission of different diffracted orders as a function of the HM spiral wavevector $q$. $\theta_{\texttt{in}}=0.2$, $\phi_{\texttt{in}}=0.5$ in radian. In panels (a) to (c), line color differentiates the HM exchange coupling strength $J$. In panel (b), solid and dotted lines are for the order -1 and 1 transmission coefficients respectively.}
\end{figure}

\begin{figure}[h]
\includegraphics[height=10cm, width=14cm]{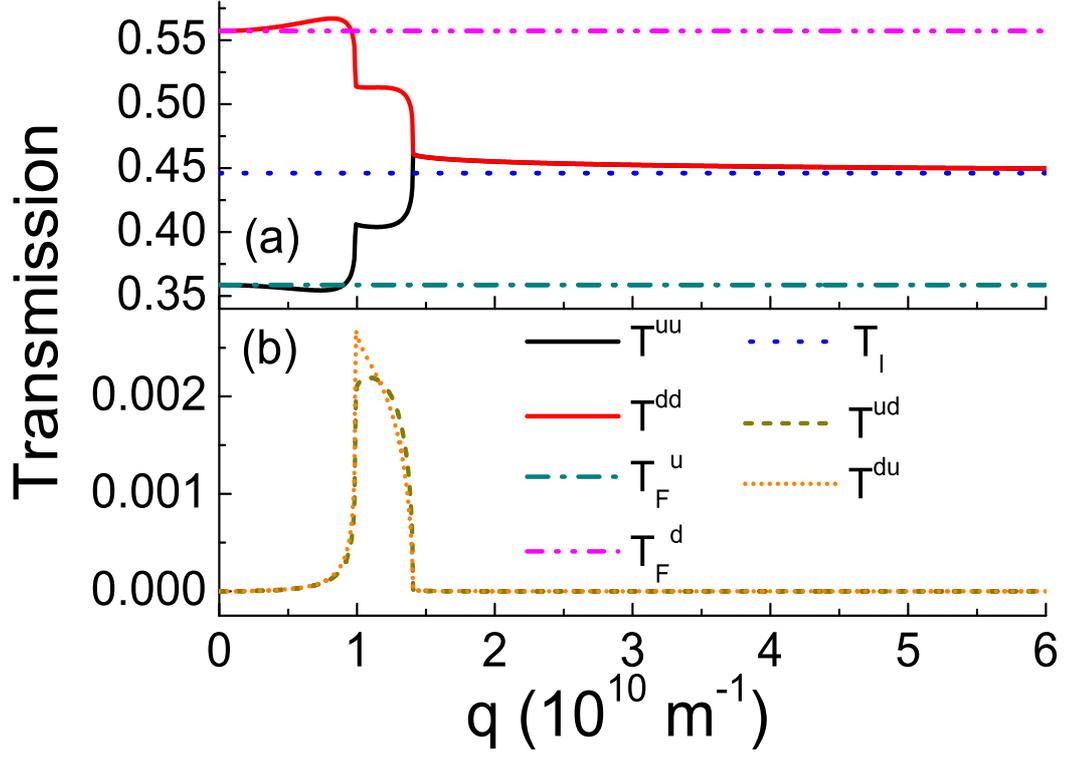}
\caption{Total spin-dependent transmission as a function of the HM spiral wavevector $q$. The HM exchange coupling strength $J=100$ meV. The incident angles are the same as Fig. 2. For reference, the spin-up and down transmissions of a ferromagnet barrier are given in green dash-dot and pink dash-dot-dot lines, respectively. Also, the transmission of a normal insulator barrier is in a blue dotted line. }
\end{figure}

\begin{figure}[h]
\includegraphics[height=10cm, width=14cm]{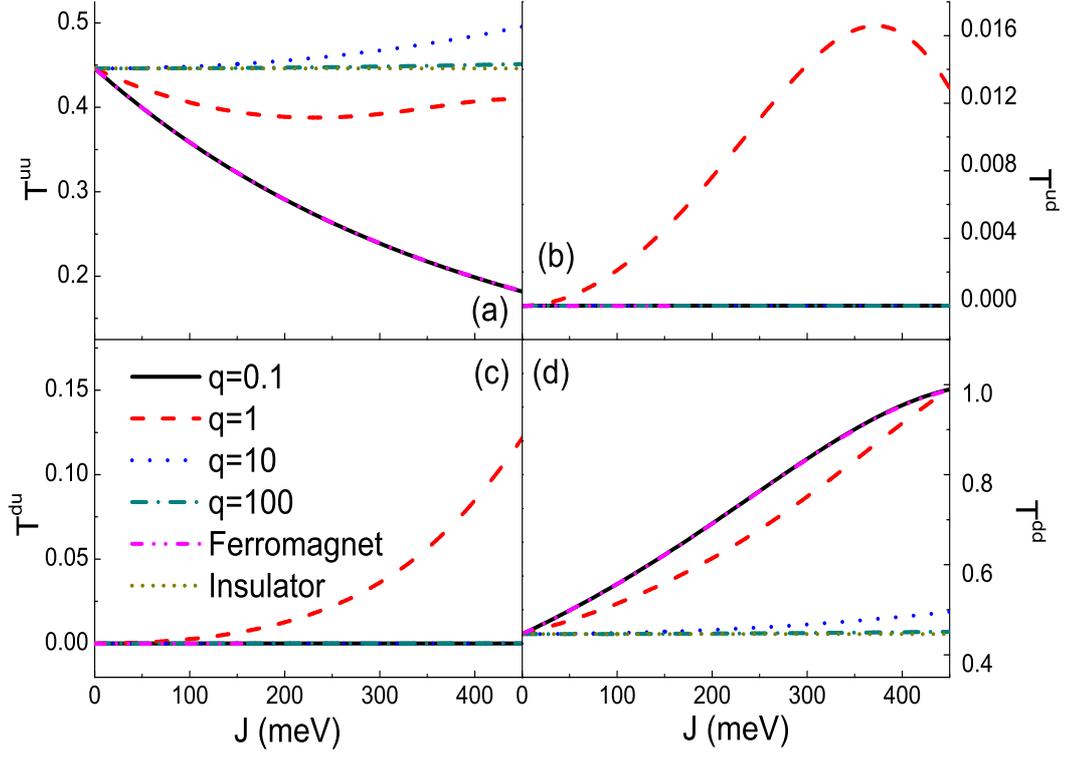}
\caption{Total spin-dependent transmission as a function of the HM exchange coupling strength $J$ for different HM spiral wavevector $q$. The incident angles are the same as Fig. 2. Transmissions of ferromagnet and normal insulator barriers are also given for reference. }
\end{figure}

\begin{figure}[h]
\includegraphics[height=10cm, width=14cm]{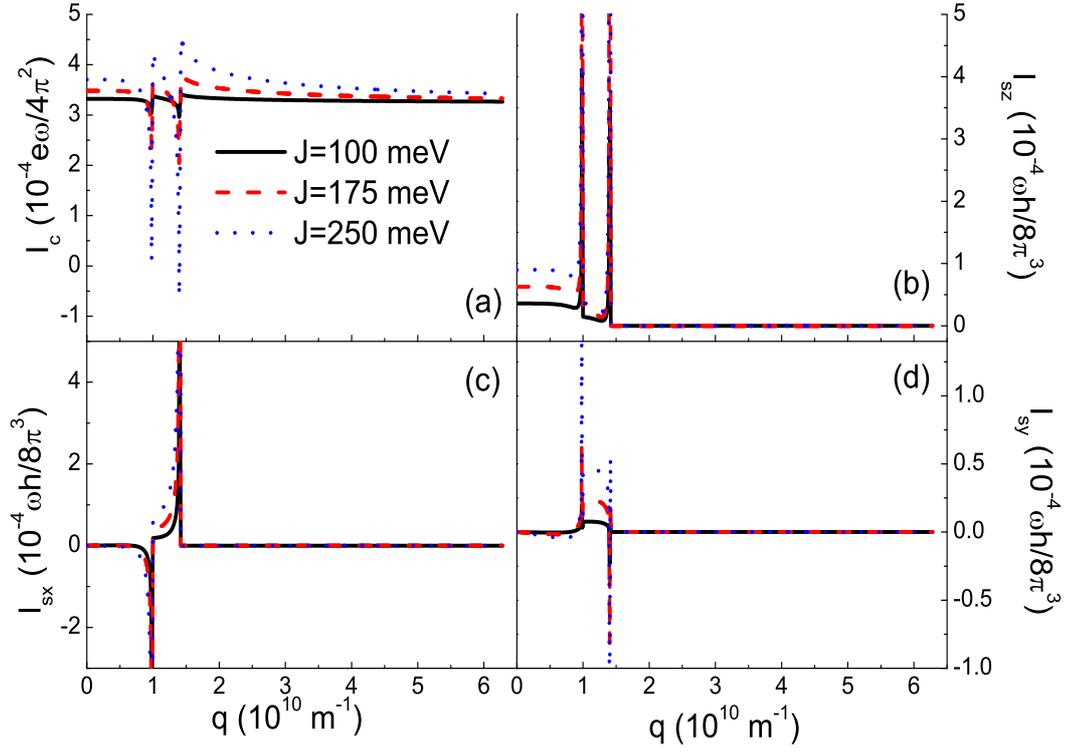}
\caption{Pumped charge and spin currents as a function of the HM spiral wavevector $q$ in a certain transport direction. Results for different HM exchange coupling strength $J$ are given. The incident angles are the same as Fig. 2. }
\end{figure}

\begin{figure}[h]
\includegraphics[height=10cm, width=14cm]{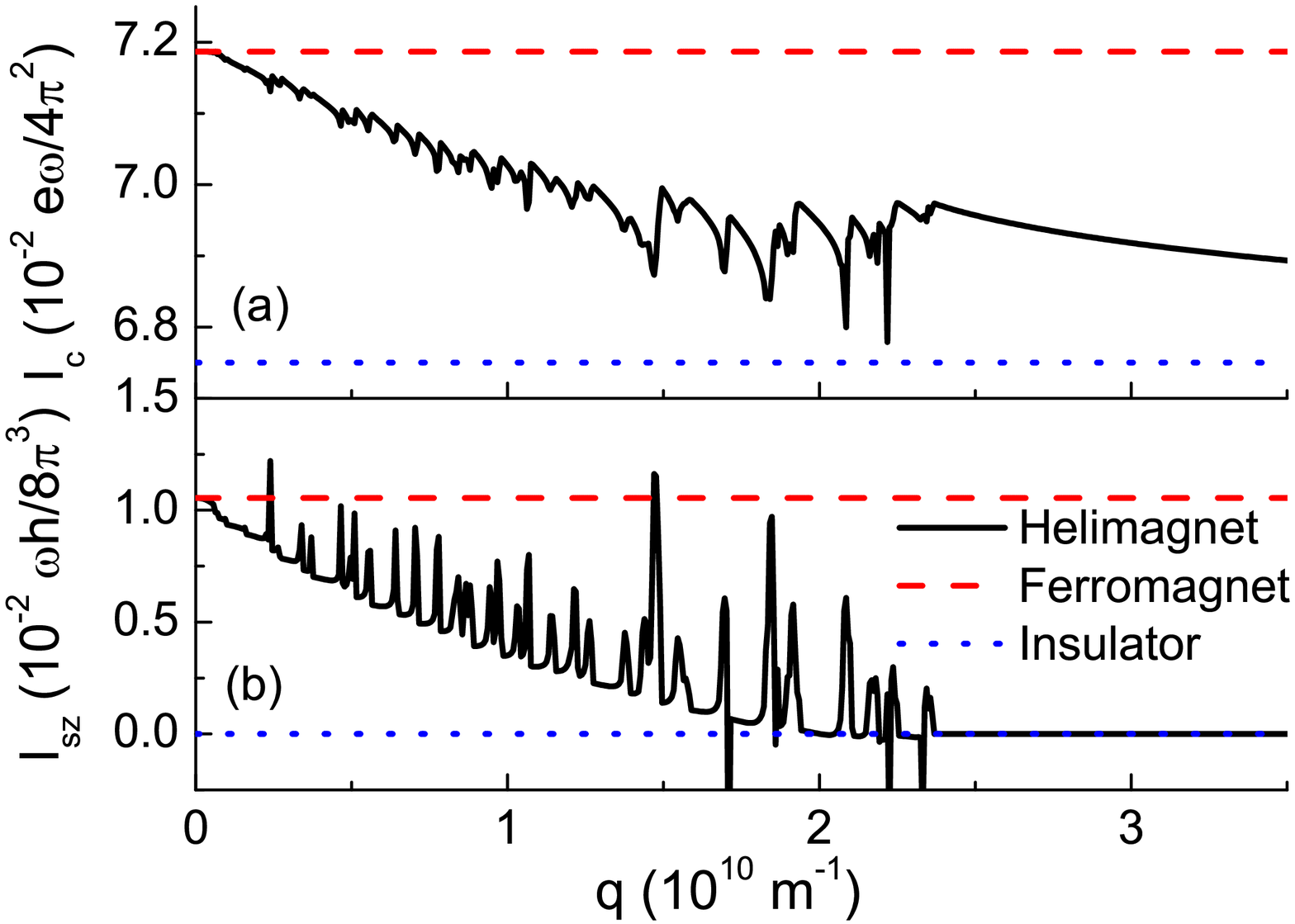}
\caption{Angular averaged pumped charge and $z$-component spin currents as a function of the HM spiral wavevector $q$. The HM exchange coupling strength $J$ is set to be 100 meV. }
\end{figure}

\clearpage

\clearpage


\clearpage




\begin{references}

\bibitem{Ref1} D. J. Thouless, Phys. Rev. B {\bf 27}, 6083 (1983).

\bibitem{Ref2} M. B\"uttiker, H. Thomas, and A. Pr\^etre, Z. Phys. B \textbf{94}, 133
(1994); Phys. Rev. Lett. \textbf{70}, 4114 (1993).

\bibitem{Ref14} M. Switkes, C. M. Marcus, K. Campman, and A. C. Gossard,
Science {\bf 283}, 1905 (1999).

\bibitem{Ref5} M. Moskalets and M. B\"{u}ttiker, Phys. Rev. B {\bf 66},
035306 (2002).

\bibitem{Ref3} P. W. Brouwer, Phys. Rev. B {\bf 58}, R10135 (1998).

\bibitem{Ref4} D. Xiao, M.-C. Chang, and Q. Niu, Rev. Mod. Phys. {\bf 82}, 1959 (2010).

\bibitem{Ref53} M. Alos-Palop, R. P. Tiwari, and M. Blaauboer, arXiv:1305.1512.

\bibitem{Ref18} R. Zhu and H. Chen, Appl. Phys. Lett. {\bf 95}, 122111
(2009).

\bibitem{Ref19} E. Prada, P. San-Jose, and H. Schomerus, Phys. Rev. B {\bf 80}, 245414 (2009).

\bibitem{Ref54} Y. -C. Xiao, W. -Y. Deng, W. -J. Deng, R. Zhu, and R. -Q. Wang, Phys. Lett. A \textbf{377}, 817 (2013).

\bibitem{Ref35} M. Alos-Palop, R. P. Tiwari, and M. Blaauboer, New J. Phys. {\bf 14}, 113003 (2012).

\bibitem{Ref20} R. Citro, F. Romeo, and N. Andrei, Phys. Rev. B {\bf 84}, 161301(R) (2011).

\bibitem{Ref22} Y. Tserkovnyak, A. Brataas, G. E. W. Bauer, and
B. I. Halperin, Rev. Mod. Phys. {\bf 77}, 1375 (2005).

\bibitem{Ref26} M. V. Costache, M. Sladkov, S. M. Watts, C. H. van der Wal, and B. J. van Wees, Phys. Rev. Lett. {\bf 97}, 216603 (2006).

\bibitem{Ref55} R. Benjamin and C. Benjamin, Phys. Rev. B \textbf{69}, 085318 (2004).

\bibitem{Ref56} R. Citro and F. Romeo, Phys. Rev. B \textbf{73}, 233304 (2006).

\bibitem{Ref57} F. Romeo and R. Citro, Eur. Phys. J. B \textbf{50}, 483 (2006).

\bibitem{Ref58} J. Splettstoesser, M. Governale, and J. K\"{o}nig, Phys. Rev. B \textbf{77}, 195320 (2008).

\bibitem{Ref39} H. Katsura, S. Onoda, J. H. Han, and N. Nagaosa, Phys. Rev. Lett. {\bf 101}, 187207 (2008).

\bibitem{Ref40} C. Jia and J. Berakdar, Phys. Rev. B \textbf{81}, 052406
(2010).

\bibitem{Ref42} A. Manchon, N. Ryzhanova, A. Vedyayev, and B. Dieny, J. Appl. Phys. \textbf{103}, 07A721 (2008);

\bibitem{Ref43} R. Zhu, arXiv:1204.6095.

\bibitem{Ref51} C. Jia and J. Berakdar, Appl. Phys. Lett \textbf{95}, 012105 (2009).

\bibitem{Ref52} R. Zhu, J. Phys.: Condens. Matter \textbf{25}, 036001 (2013).

\bibitem{Ref45} Y. Tserkovnyak and A. Brataas, Phys. Rev. B {\bf 71}, 052406 (2005).

\bibitem{Ref44} N. Kanazawa, Y. Onose, T. Arima,
D. Okuyama, K. Ohoyama, S. Wakimoto, K. Kakurai, S. Ishiwata, and Y.
Tokura, Phys. Rev. Lett. \textbf{106}, 156603 (2011).



\end{references}
\end{document}